\documentclass[aps,prl,twocolumn,superscriptaddress]{revtex4-2}
\usepackage{graphicx,soul}
\usepackage{amsmath}
\usepackage{xcolor}
\usepackage{hyperref}
\usepackage{textcomp}

\begin{document}

\title{Demonstration of unpartible entanglement}

\author{Philip Held}
\email{philip.held@uni-paderborn.de}
\affiliation{Paderborn University, Integrated Quantum Optics, Institute for Photonic Quantum Systems (PhoQS), Warburger Str. 100, 33098 Paderborn, Germany}

\author{Laura Ares}
\affiliation{Paderborn University, Theoretical Quantum Science, Institute for Photonic Quantum Systems (PhoQS), Warburger Str. 100, 33098 Paderborn, Germany}

\author{Federico Pegoraro}
\affiliation{Paderborn University, Integrated Quantum Optics, Institute for Photonic Quantum Systems (PhoQS), Warburger Str. 100, 33098 Paderborn, Germany}

\author{Jonas Lammers}
\affiliation{Paderborn University, Integrated Quantum Optics, Institute for Photonic Quantum Systems (PhoQS), Warburger Str. 100, 33098 Paderborn, Germany}

\author{Benjamin Brecht}
\affiliation{Paderborn University, Integrated Quantum Optics, Institute for Photonic Quantum Systems (PhoQS), Warburger Str. 100, 33098 Paderborn, Germany}

\author{Jan Sperling}
\affiliation{Paderborn University, Theoretical Quantum Science, Institute for Photonic Quantum Systems (PhoQS), Warburger Str. 100, 33098 Paderborn, Germany}

\author{Christine Silberhorn}
\affiliation{Paderborn University, Integrated Quantum Optics, Institute for Photonic Quantum Systems (PhoQS), Warburger Str. 100, 33098 Paderborn, Germany}

\begin{abstract}
We report on the first experimental verification of mode-independent entanglement.
Commonly, the entanglement of a state is firmly based on pre-defined parties that are correlated, and the state might be disentangled when the definition of the parties is changed.
Exceeding this party-dependent concept, we realize a type of quantum entanglement that persists even if the parties, in our case modes, are transformed.
This safeguards the performance of entanglement in real-world applications, such as quantum communication settings involving noise and untrusted parties.
For the state generation, we present an experimental scheme based on a fully reconfigurable temporally multiplexed interferometer with measurement-induced nonlinearities, which generates heralded two-photon states in two modes that are entangled for all choices of orthonormal mode basis.
For the certification process, we utilize a tailored quantum-state tomography, achieving fidelities that validate the presence of mode-independent entanglement as a resilient and operationally advantageous quantum correlation.
\end{abstract}

\maketitle

\paragraph*{Introduction.---}
Entanglement \cite{neumann,RevEnt} constitutes a key resource for quantum technologies, such as communication \cite{E91}, sensing \cite{FreqEnt}, and computing \cite{QuComp}.
As this quantum feature emerges from the non-separability of parties, it is, by construction, dependent on the choice of the subsystems that constitute a party.
Diverse notions of entanglement emerge based on the physical property that defines each party.
One example is particle entanglement, as in the case of EPR states \cite{EPR}, a prime resource for quantum teleportation \cite{QuTele} and non-locality tests \cite{NonLocal}.
Further examples show up when the parties considered are a system’s different degrees of freedom, leading to, for example, entanglement within single \cite{1PartEnt} or multiple particles \cite{2PartEnt}, between electromagnetic fields \cite{Enk}, and between the internal and external degrees of freedom of a quantum walker \cite{EntQW}.
In particular, entanglement between optical modes \cite{ModeEnt} has been shown to be a powerful resource for quantum technologies \cite{N00Nmeas, singlePhoComm, singlePho32Time}.

In mode entanglement, the identification of the subsystems---who is Alice and who is Bob---depends on the chosen mode basis \cite{ModeBasis}.
Thus, the decomposition into subsystems can be readily modified by transformations across modes.
As a consequence, such ``nonlocal" transformations can profoundly affect whether a given state is entangled or separable \cite{Enk}.
This feature poses a challenge for quantum protocols operating over non–mode-preserving channels \cite{RobustEnt}, motivating the search for states that remain entangled under arbitrary mode representations, and leading to the definition of mode-independent quantum entanglement (MIQE) \cite{theo_orig}.
This robust form of quantum correlations prevents even partial separability of the states under any linear unitary transformation of the bosonic creation operators, i.e., the choice of the mode basis.
Note that the resulting states are thus not maximally entangled in any basis; however, they remain entangled even in the absence of a pre-defined mode basis \cite{theo_orig}.

While the controlled generation of various entangled states has seen significant advances, the reliable certification of entanglement under realistic conditions remains a central challenge \cite{EntHard}.
Beyond the intrinsic NP-hardness of deciding separability \cite{LG03,SG10}, imperfections in practical implementations, such as noise and finite statistics, severely limit the reliable observation of these correlations \cite{GT2009}.
For this reason, entanglement validation increasingly relies on witness-based criteria \cite{EntWit2,SV13}, which are experiment-tailored bounds.
These witnesses are sufficient but not necessary criteria for entanglement \cite{EntWit1}, yet directly applicable.
Experimentally, several strategies are used to measure the quantities required for witnessing entanglement.
Examples are quantum state tomography \cite{FullTomo}, reduced schemes like shadow tomography \cite{Shadow}, and quantum reservoir computing \cite{QuResComp}.

In this letter, we report on the first certification of mode-independent entanglement \cite{theo_orig}.
To this end, we generate different states within the most fundamental family of states that present this quantum correlation, i.e., involving the minimal number of modes and particles, in a heralded scheme.
While this concept applies to all bosonic systems, we implement it in a photonic architecture, a promising platform for a variety of quantum technologies; for a review, see \cite{PhoQuTec}.        
The different states are generated using a fully reconfigurable temporally multiplexed interferometer (TMI) \cite{CNOT} with measurement-induced non-linearities.
A tailored quantum-state tomography is employed to analyze the generated states.
We witness the presence of MIQE in several of the reconstructed states with up to six standard deviations. 
Furthermore, we explicitly show the persistence of this quantum correlation under mode transformations for a generated state, opening a path toward realistic mode-entanglement-based quantum technologies.

\paragraph*{Theoretical framework.---} 
In order to describe MIQE, we consider two orthogonal optical modes $a$ and $b$ represented by the bosonic creation operators $\hat{a}^\dagger$ and $\hat{b}^\dagger$, respectively.
Transformations between modes are performed by unitary operators, 
\begin{equation}
\label{eq:Transformation}
    \hat U(\theta,\phi)=
    \begin{pmatrix}
        \cos(\theta/2) & e^{i\phi}\sin(\theta/2) \\
        - e^{-i\phi}\sin(\theta/2) & \cos(\theta/2)
    \end{pmatrix}.
\end{equation}
Furthermore, let us recall that a pure state $|\psi\rangle$ with a fixed total photon number $n+m$ is entangled when $|\psi\rangle\neq|n\rangle_a\otimes|m\rangle_b=|n,m\rangle_{a,b}$, where $n$ and $m$ represent the number of photons in each respective mode.

Before presenting the family of states under study, exhibiting the MIQE property, we contrast two states broadly utilized in photonics to build an intuition for this quantum feature.
Combining two indistinguishable photons, $|1,1\rangle_{a,b}$, one on each input port of a balanced beam splitter (BS), yields a state of the form $(|2,0\rangle_{a,b} - |0,2\rangle_{a,b})/\sqrt{2}$ because of Hong-Ou-Mandel interference (HOMI) \cite{N00N,HOM}.
As this two-particle N00N state possesses mode entanglement that has been generated from a separable state by linear unitary operations, it is clearly not resilient against changes in the mode representation.
Another example worth to mention are Bell states,  $(|1,0,1,0\rangle_{1a,1b,2a,2b} - |0,1,0,1\rangle_{1a,1b,2a,2b})/\sqrt{2}$, where subscripts 1 and 2 refer to the two particles involved.
Experimental validation of MIQE for these states is not possible so far since no criteria for certifying it in this class of states have been developed for four-mode states. 

\begin{figure}[t]
	\centering
	\includegraphics[width=1\linewidth]{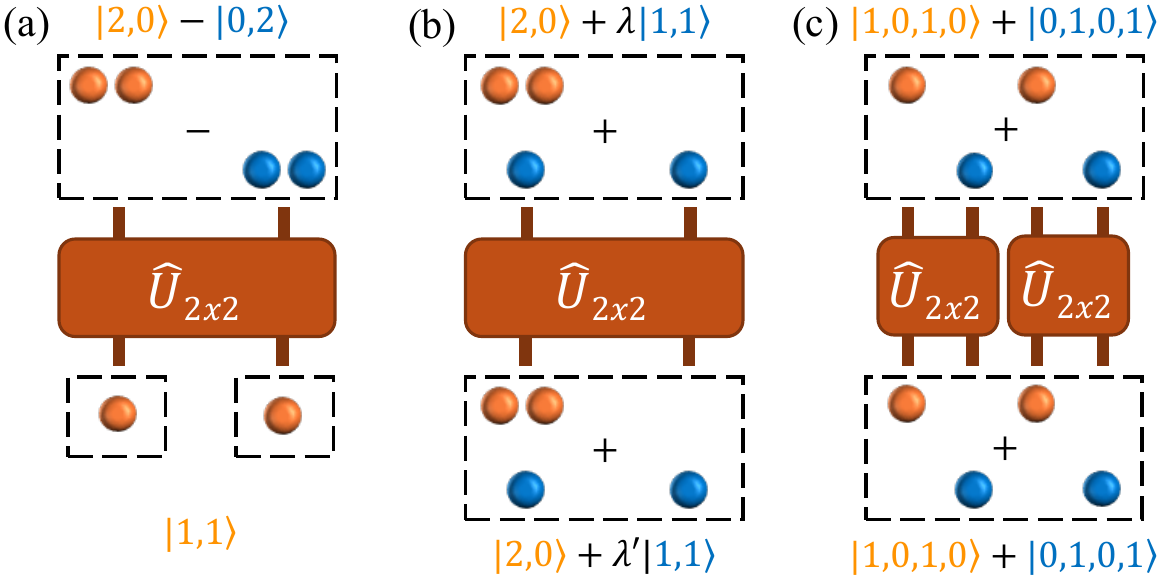}
	\caption{Representation of different unnormalized two-particle entangled states:
	(a) N00N state for $N=2$, with mode entanglement that can be erased by a linear mode transformation.
	(b) MIQE state, with mode entanglement that survives under any linear mode transformation.
	(c) Bell state, a particle-entangled state for which no criteria to certify MIQE exist.}
	\label{fig:concept}
\end{figure}

We now present the family of states involving the minimum number of modes and photons that carry the MIQE property \cite{theo_orig}, dubbed MIQE states throughout the remainder of this work.
Such states are generated when two photons are excited in two non-parallel and non-orthogonal modes, i.e., modes of the form $\hat{a}^\dagger$ and $\hat{a}^\dagger + \lambda \hat{b}^\dagger$, respectively, with $\lambda$ being a nonzero complex parameter.
This construction leads to the family of states
\begin{equation}
\label{eq:MIQEstate}
	|\Psi(\lambda)\rangle = \frac{\sqrt{2} |2,0\rangle_{a,b} + \lambda |1,1\rangle_{a,b}}{\sqrt{2 + |\lambda|^2}}.
\end{equation}
A schematic comparison between N00N, Bell, and the family of MIQE states regarding the persistence of entanglement under mode transformations is depicted in Fig. \ref{fig:concept}.

To certify mode-independent entanglement, we employ the measurable criterion derived in Ref. \cite{theo_orig}.
It is based on  the idea of entanglement witnesses:
Only entangled states $\hat{\rho}$ can produce expectation values of a certain observable $\hat{L}$ exceeding the maximal value attainable by separable states $\hat{\sigma}$ \cite{sigma,SV13}, thus $\mathrm{tr}(\hat{\rho}\hat{L})>\sup_{\hat\sigma} \mathrm{tr}(\hat\sigma\hat L)$.
Since the observable utilized is the projector onto the target state, its expectation value  corresponds to a quantum state fidelity (QSF)
\begin{equation}
    \mathcal{F}=\mathrm{tr}(\hat\rho|\Psi(\lambda)\rangle\langle\Psi(\lambda)|).
\end{equation}
Accordingly, the bound constitutes the maximal overlap between Eq. (\ref{eq:MIQEstate}) and a separable state. 
In order to witness entanglement in different base representations, one applies a mode transformation to the target state, thus $\mathcal{F}_{\hat{U}}=\textrm{sup}_{\hat{\sigma}} \left[\textrm{tr}\left(\hat{\sigma} \hat{U}|\Psi(\lambda)\rangle\langle\Psi(\lambda)|\hat{U}^\dagger\right)\right]$.
The bound for mode-independent entanglement is then the maximum over all mode transformations $\mathcal{F}_\mathrm{bound}=\textrm{sup}_{\hat{U}}\left[\mathcal{F}_ {\hat{U}}\right(\lambda)]$.
With Eqs. (\ref{eq:Transformation}) and (\ref{eq:MIQEstate}), this bound reads as \cite{theo_orig}
\begin{equation}
	\label{eq:F_bound} 
    \mathcal{F}_\mathrm{bound}(\lambda) = \textrm{max} \left\{\frac{1}{2}+\frac{\sqrt{1+|\lambda|^2}}{2+|\lambda|^2}, 1 - \frac{1}{+|\lambda|^2}\right\}.
\end{equation}

Following the witness-like procedure, if one can measure QSFs surpassing the bound in Eq. (\ref{eq:F_bound}), the presence of mode-independent entanglement in the corresponding state is certified. 
Note that because this criterion is sufficient but not necessary, it can only verify the presence of MIQE. 
Finally, since $\mathcal{F}_\mathrm{bound}(\lambda)>0.85$, witnessing MIQE implies the measurement of QSFs over $85\%$, which requires a high-quality state generation and a neat state reconstruction.

\begin{figure*}[t]
	\centering
	\includegraphics[width=1\linewidth]{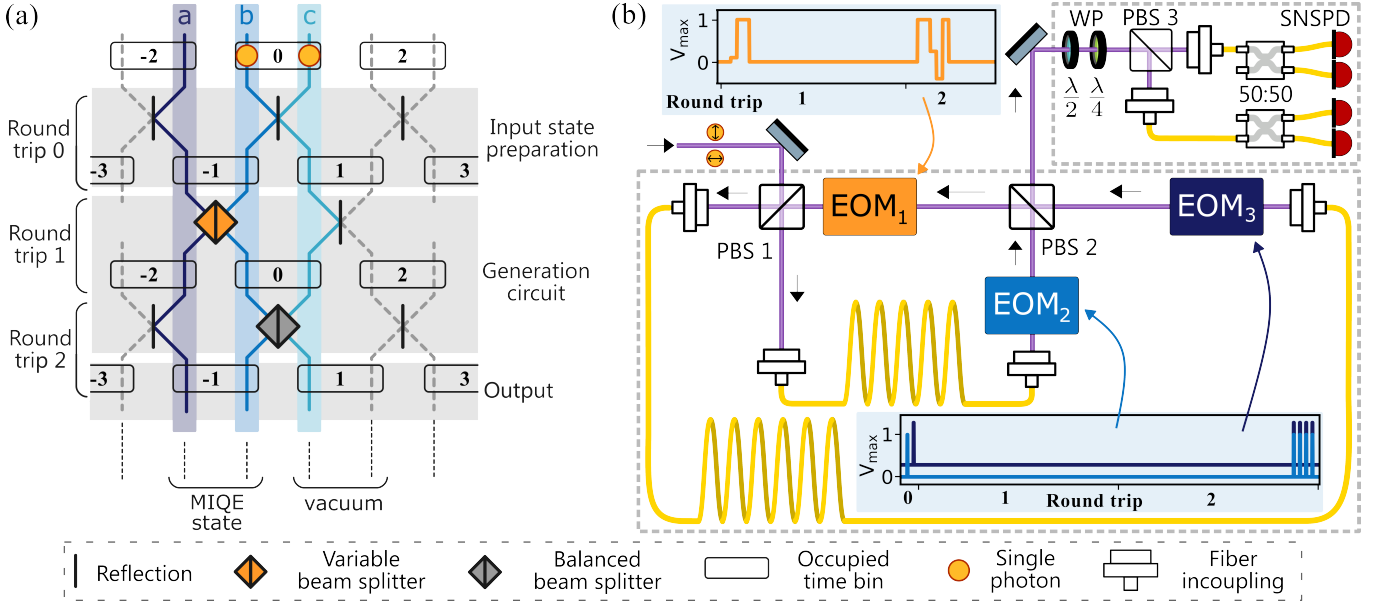}
	\caption{(a) Schematic of the operation principle of the photonic circuit.
	(b) Schematic of the TMI implementing the photonic circuit and the polarization tomography.
    Dynamic control is enabled by EOMs, for which exemplary control sequences are shown; for clarity, vertical offsets are added.}
	\label{fig:setup}
\end{figure*}

\paragraph*{Operating principle.---}
The schematic of the photonic circuit utilized to generate MIQE states is depicted in Fig. \ref{fig:setup}(a).
The rows in the sketch denote the layers of the circuit, implemented by individual round trips (RT) through the TMI, while the columns are temporal pulse positions, refereed as time bins (TB), within each RT.
Note that due to the structure of the TMI in a particular RT, only even or odd TBs are occupied, indicated by rounded rectangles, containing two modes.

The generation circuit itself requires three modes, here referred to as $a$, $b$, and $c$, and a depth of two layers.
A MIQE state is generated across the output modes $a$ and $b$ upon heralding on no photon populating output mode $c$.
Initially, the modes $b$ and $c$ are populated at TB 0 with one photon each, thus $|\psi_\textrm{in}\rangle = |0,1,1\rangle_{a,b,c}.$
In the input state preparation, they are redirected to the TBs $-1$ and 1 for further processing, while remaining in the modes $b$ and $c$.
This is a consequence of the change in the order of occupied TBs from step to step in our TMI.

In the first layer of the generation circuit, modes $a$ and $b$ are mixed on a variable BS (orange).
This controls the weighting parameter $\lambda$ in Eq. (\ref{eq:MIQEstate}) by varying the reflectivity $R$ as $|\lambda|=\sqrt{2}/\tan\left[\arcsin\left(\sqrt{R}\right)\right]$.
In this manner, we can generate the complete family of two-photon-two-mode MIQE states by only reconfiguring this single BS.
This results in the state $|\Psi_\textrm{1}\rangle =  \frac{-\sqrt{2} |0,1\rangle_{a,b} + \lambda |1,0\rangle_{a,b}}{\sqrt{2 + |\lambda|^2}} \otimes |1\rangle_c$.
In the second layer, a balanced BS between modes $b$ and $c$ enables photon bunching, i.e., HOMI.
Thus, we obtain the state
\begin{align}
\label{eq:final}
|\Psi_\textrm{2}\rangle =& \frac{\sqrt{2} |0,2\rangle_{a,b} + \lambda |1,1\rangle_{a,b}}{\sqrt{2} \sqrt{2 + |\lambda|^2}} \otimes |0\rangle_c \nonumber\\
&+ \frac{\lambda |1,0\rangle_{a,b}}{\sqrt{2} \sqrt{2 + |\lambda|^2}}  \otimes |1\rangle_c- \frac{\sqrt{2} |0,0\rangle_{a,b}}{\sqrt{2} \sqrt{2 + |\lambda|^2}}  \otimes |2\rangle_c.
\end{align}

Finally, observing no photon in TB 1 corresponds to projecting mode $c$ into $|0\rangle_c$.
Thus, we herald the desired MIQE state in the form of Eq. (\ref{eq:MIQEstate}) in the modes $a$ and $b$ in TB $-1$, with a circuit's success rate of 50\%.

\paragraph*{Experimental setup.---}
For the circuit, two indistinguishable photons are required as the resource state, which are generated in a degenerate type-II parametric down-conversion process.
A dispersion-engineered periodically poled potassium titanyl phosphate waveguide is pumped by a Ti:Sapph mode-locked oscillator at 772.5 nm.
This produces photon pairs at 1545 nm, with one horizontally (H) and one vertically (V) polarized photon that exhibit a HOMI visibility of $V = (98 \pm1)$\%; for a detailed description of the source, see \cite{ConditionedQWs}.
The source is operated in the low-gain regime with a mean photon number below 0.01, reducing noise caused by higher photon-number contributions.

A TMI with hybrid temporal and polarization encoding is chosen for implementing the photonic circuit, as temporally multiplexed architectures have the advantages of resource efficiency \cite{1stQW}, scalability \cite{Xanadu_TM}, and intrinsic interferometric stability \cite{NoiseStabil}.
Here, a single BS acting on polarization is reused sequentially over time to implement the BSs that are arranged spatially in the schematic circuit.
This is achieved by routing the BS's outputs back to its inputs via optical delay lines, implementing the RTs.
Thus, the two modes per time bin are encoded in H and V  polarization.
The delay lines introduce different delays for the two polarizations, leading to the TB shift.
The BS's reflectivity needs to change fast enough to implement the individual BS operations in time.
Employing free-space electro-optical modulators (EOMs), switching times of the polarization in the range of $100$ ns are performed, while keeping the loss rate low.

Figure \ref{fig:setup}(b) shows an illustration of the TMI.
EOM 1 implements the fast switching variable BS operation.
Then, polarizing BS (PBS) 1 separates both polarizations such that the polarization-dependent delays can be applied using optical fibers of 1115 m (H) and 1080 m (V), leading to an RT time and TB separation of 5.3 $\mu$s and 172 ns, respectively.
The two spatial paths are recombined at PBS 2, concluding a RT.
To route the photons in and out of the TMI, two additional EOMs (2 and 3) are used.
These control the photons' polarization, directing them either to EOM 1, for an additional RT, or to the detection.

To validate the successful generation of MIQE states, a polarization state tomography is performed \cite{Tomo}.
This comprises a half- and quarter-wave plate followed by PBS 3 and photon-number resolved (PNR) detection.
This is implemented by a 50:50 BS and two superconducting nanowire single photon detectors (SNSPDs) per mode.

We identify three well-characterized imperfections in our detection setup.
First, one SNSPD features 70\% of the efficiency of the other three.
Second, due to the implementation of the PNR, the recorded number of two-photon events in the same mode is reduced by a factor of two.
Third, the PBS exhibits 1\% leakage of H polarized light.
These imperfections are accounted for in the reconstruction process by updating the recorded clicks accordingly. 
The details of the state reconstruction from the measured click statistics are given in the End Matter.

\begin{figure}[t]
	\centering
	\includegraphics[width=1\linewidth]{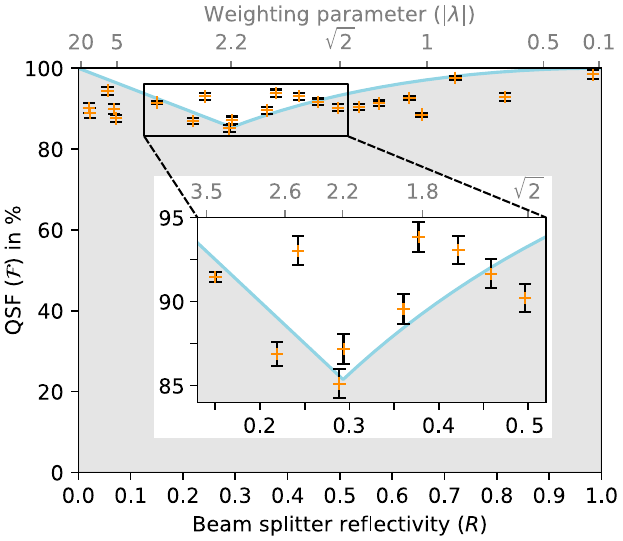}
	\caption{QSFs for the reconstructed density operators (orange pluses) with an error margin of one standard deviation in black against the
    bound for certifying MIQE $\mathcal{F}_\mathrm{bound}(\lambda)$ (blue line).}
	\label{fig:states}
\end{figure}

\paragraph*{Results.---}
We generate the two-photon, two-mode states in Eq. (\ref{eq:MIQEstate}), with $\lambda\in[0.18,9.9]$, by adjusting the reflectivity of the variable BS in the circuit. 
Our main result is presented in Fig. \ref{fig:states}, which reports the measured QSFs of the reconstructed states.
The MIQE witnessing bound, $\mathcal{F}_{\mathrm{bound}}$ [Eq. (\ref{eq:F_bound})], is indicated in blue and reaches its minimum at $|\lambda| \approx 2.2$.
In the scanned regime, we successfully certify MIQE with a statistical significance of up to six standard deviations (for $|\lambda| = 2.50$).
This unambiguously demonstrates the verification of MIQE.

As $|\lambda|$ approaches 0 and $\infty$, the QSF necessary for certifying MIQE tends to 100\%. 
In these limits, the states in Eq. (\ref{eq:MIQEstate}) become separable, and MIQE disappears.
We observe an average QSF of $ (91.1 \pm 3.3)$\%, with a highly homogeneous distribution across the explored parameter range.
This confirms a consistently good performance of the generation circuit.

\begin{figure}[t]
	\centering
	\includegraphics[width=1\linewidth]{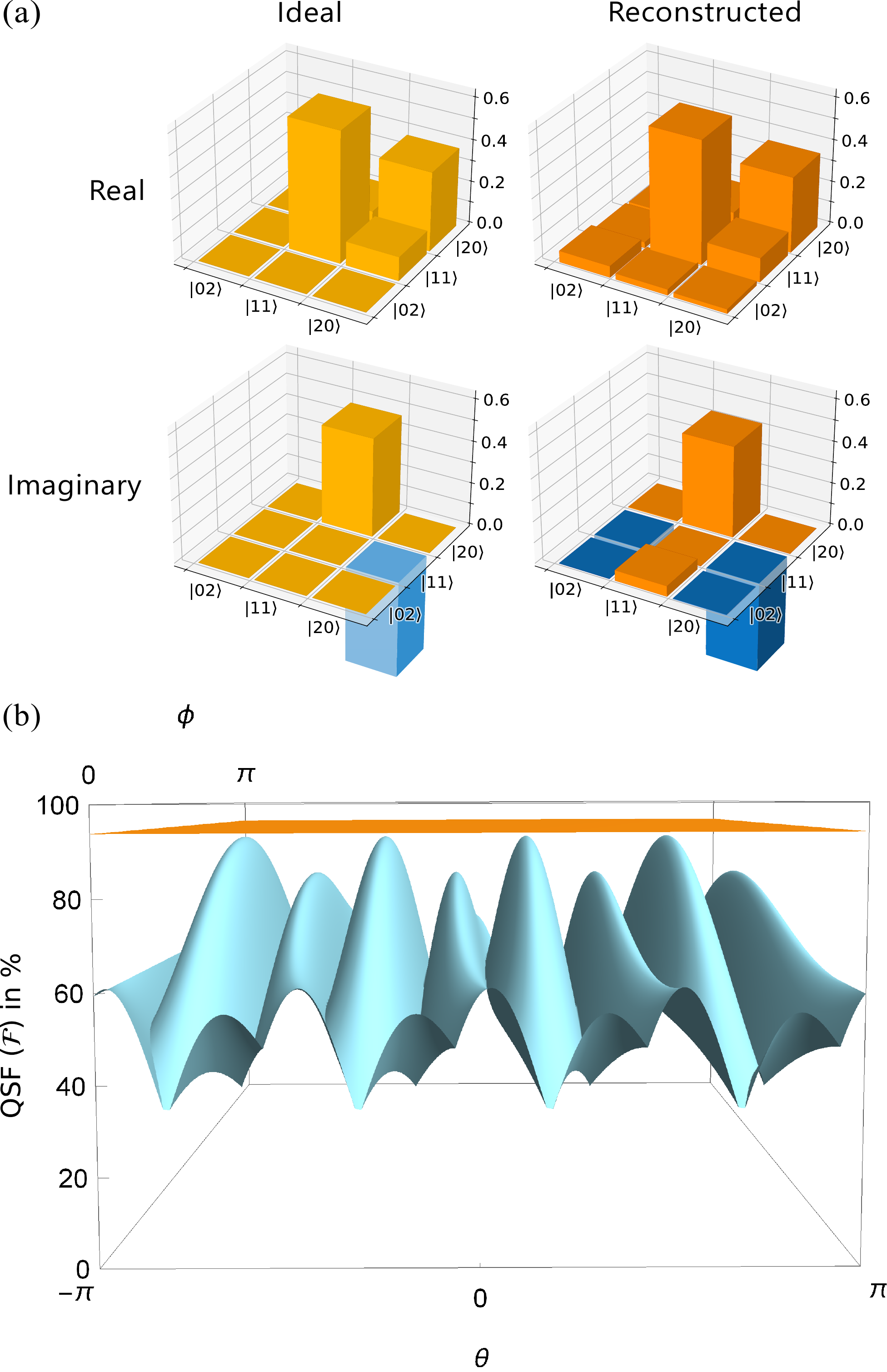}
	\caption{(a) Real and imaginary part of the density operators for $|\lambda| = 1.82$.
	The experimentally reconstructed and ideal states feature a QSF of $\mathcal{F}=(93.8 \pm 0.9)$\%.
	(b) Entanglement witness $\mathcal{F}_{\hat{U}}$ (blue) for $|\lambda| = 1.82$ for any linear optical transformation of the mode basis.
	MIQE is confirmed as $\mathcal{F}$ (orange) surpasses $\mathcal{F}_{\hat{U}}$ over mode transformations, cf. Eq. (\ref{eq:Transformation}). }
	\label{fig:rho}
\end{figure}

Figure \ref{fig:rho}(a) shows the reconstructed density operator for $|\lambda| = 1.82$, together with the ideal, theoretically predicted one.
The main deviation corresponds to the $|0,2\rangle$ components, which do not vanish due to the small amount of higher photon-number contributions from the source, imperfect projections in the tomography, and the minimal addition of white noise applied to enforce positivity.
Despite these effects, the reconstructed state achieves a QSF of $(93.8 \pm 0.9)$\%, displaying a violation of the witness bound by 5.3 standard deviations.

To explicitly show the MIQE property of the state in Fig. \ref{fig:rho}(a), we present the witness for entanglement corresponding to each mode basis $\mathcal{F}_{\hat{U}}$---blue surface in Fig. \ref{fig:rho}(b)---, where $\hat U$ is parametrized as in Eq. (\ref{eq:Transformation}).
It varies significantly with the basis transformations $\hat U(\theta,\phi)$.
A QSF around 63\% would be enough for certifying entanglement in the generation basis ($\theta=\phi=0$).
Nonetheless, this is insufficient for entanglement certification if the basis suffers small rotations (e.g., $\theta/2\approx\pi/20$) and phase shifts.
Thus, to ensure entanglement after any linear mode transformation, the reconstructed QSF needs to exceed the global maximum of $\mathcal{F}_{\hat{U}}$, i.e., $\mathcal{F}_{\mathrm{bound}}$.
This is achieved by our experimental value (orange plain).

\paragraph*{Summary.---}
We certify with high statistical significance the experimental generation of mode-independent quantum entanglement.
We present a high-performance scheme for generating heralded two-photon, two-mode MIQE states based on a temporally multiplexed interferometer.
The resulting states are encoded in polarization, ensuring compatibility with various applications.
Thus, our setup constitutes a reliable source of MIQE, beneficial, for example, for applications in quantum metrology \cite{sensing}.
We further utilize a tailored tomography scheme that enables high-fidelity reconstruction of the states, allowing the experimental certification of MIQE and the explicit demonstration of entanglement persistence under arbitrary unitary mode transformations.
Thus, our results demonstrate the suitability of this resilient form of mode entanglement for quantum information processing in realistic, non-mode-preserving scenarios.

\paragraph*{Acknowledgments.---}
P.H., F.P., J.L., B.B, and C.S. acknowledge financial support by the European Commission through the Horizon Europe project EPIQUE (Grant No. 101135288).
The authors received funding from the German Federal Ministry of Research, Technology and Space (BMFTR) within the PhoQuant project (Grant No. 13N16103).
L.A. and J.S. acknowledge funding through the QuantERA project QuCABOoSE.

\paragraph*{Data availability.---}
The code and data for reconstructing the density operators are openly available \cite{code}.

\appendix*
\section{End Matter}

\paragraph*{State reconstruction.---}
For a tomographically overcomplete measurement, we utilize 8 different measurement bases.
These offer an enhanced statistical significance for a high-enough quality state reconstruction, keeping the demands of the measurement process to a minimum. 
For each measurement $l=[\mathrm{\theta_{HWP}, \theta_{QWP}}]$, see \cite{footnote} for values, we restrict the recorded counts $C$ to the two-photon subspace and compute the probabilities of finding $k$ photons in mode $b$, $p_k(l)=\frac{C_{k,2-k}}{ C_{0,2}+ C_{1,1}+ C_{2,0}}$.
In terms of the measurement operators, these read $p_k(l) =\mathrm{tr}\left[\hat\rho\hat\Pi_{k}(l)\right]$, where each projector $\hat\Pi_{k}(l)$ is computed from the operator without wave-plate rotations, $|k,2-k\rangle\langle k,2-k|$, using the mode transformation relation in Eq. (\ref{eq:Transformation}).
The explicit relation between the rotation angles of the wave plates and the parameters of the mode transformation $\hat{U}(\theta,\phi)$ is
\begin{equation}
    \begin{bmatrix}
        \cos(2\theta_\mathrm{QWP})\sin(4\theta_\mathrm{HWP}-2\theta_\mathrm{QWP})\\
        \sin(2\theta_\mathrm{QWP})\\
        \cos(2\theta_\mathrm{QWP})\cos(4\theta_\mathrm{HWP}-2\theta_\mathrm{QWP})
    \end{bmatrix}
    =
    \begin{bmatrix}
        \sin\theta\cos\phi\\
        \sin\theta\sin\phi\\
        \cos\theta
    \end{bmatrix}.
\end{equation}

After this transformation, we encapsulate the relation between the probabilities and the density matrix, $\vec p=Q\vec\rho$, where the vectors $\vec\rho=[\rho_{i, j}]_{(i,j)}$ and $\vec p=[p_k(l)]_{(k,l)}$ are represented by a pair of indices \cite{AP25}.
The matrix $Q$ is numerically inverted by applying the Moore-Penrose inverse \cite{BH12} to obtain the entries of the density matrix, $\vec\rho=Q^{-1}\vec p$.

The uncertainties are computed via standard error computation of photon-counting data, $\sigma(p_k)=\sqrt{\frac{p_k(1-p_k)}{\sum_{k=0}^N C_{k,2-k}}}$, which is propagated via quadratic error propagation techniques, $\sigma(\cdot)=\sqrt{\sum_k \left|\frac{\partial (\cdot)}{\partial p_k}\right|^2\sigma(p_k)^2}$.    

Because of noise and finite statistics \cite{Noise1, Noise2}, insignificant negative eigenvalues (on the order of $-0.05$) appear in some reconstructed density matrices.
To ensure physicality, we add the minimal amount of white noise that renders all reconstructed states positive semidefinite.
See \cite{code} for the complete reconstruction algorithm.

\end{document}